\documentclass[aps,pra,twocolumn,floatfix,floats,showpacs,superscriptaddress,raggedbottom]{revtex4-1}
\usepackage{graphicx,latexsym}
\usepackage{dcolumn}
\usepackage{amsmath}
\usepackage{amssymb,bm}
\usepackage{braket}
\usepackage[normalem]{ulem}

\def\eq#1{Eq.\ (\ref{#1})}
\def\fig#1{Fig.\ \ref{#1}}

\usepackage{hyperref}
\hypersetup{
    pdfnewwindow=true,       
    colorlinks=true,         
    linkcolor=blue,          
    citecolor=blue,          
    filecolor=magenta,       
    urlcolor=black           
}

\begin{document}
\title{Magnetic-field effects on photon-induced quantum transport in \\ a single dot-cavity system}

\author{Nzar Rauf Abdullah}
\email{nzar.r.abdullah@gmail.com}
\affiliation{Physics Department, Faculty of Science and Science Education, School of Science, 
             University of Sulaimani, Kurdistan Region, Iraq}
\affiliation{Science Institute, University of Iceland, 
             Dunhaga 3, IS-107 Reykjavik, Iceland}
             

\author{Aziz H. Fatah}
\affiliation{Physics Department, Faculty of Science and Science Education, School of Science, 
             University of Sulaimani, Kurdistan Region, Iraq}
 
\author{Jabar M. A. Fatah}
\affiliation{Physics Department, Faculty of Science and Science Education, School of Science, 
             University of Sulaimani, Kurdistan Region, Iraq}
 

\begin{abstract}
In this study, we show how a static magnetic field can control photon-induced electron transport
through a quantum dot system coupled to a photon cavity. 
The quantum dot system is connected to two electron reservoirs and  
exposed to an external perpendicular static magnetic field.
The propagation of electrons through the system is thus influenced by the static magnetic and the dynamic photon fields. 
It is observed that the photon cavity forms photon replica states controlling electron transport
in the system. 
IF the photon field has more energy than the cyclotron energy, then the photon field is dominant
in the electron transport.
Consequently, the electron transport is enhanced due to activation of photon replica states.
By contrast, the electron transport is suppressed in the system
when the photon energy is smaller than the cyclotron energy.
\end{abstract}



\maketitle

\section{Introduction}

Quantum dot (QD) is a crucial electronic structure in the technological devices~\cite{Imamog72.210,Loss.57.1998,DiVincenzo.309.2005} 
because of it's unique properties such as zero-dimensional confinement effect~\cite{Petroff1997} and 
single electron charge effect (Coulomb blockade)~\cite{Kouwenhoven1999}. There are several methods that have been used to control 
electron motion in QD system. 
One of which is to control the energy levels and 
the electron concentration in the QD systems using a plunger gate voltage~\cite{Kouwenhoven413.2000}. 
On the other hand, by applying a photon radiation that interacts with the 
electrons in the QD arising a fascinated physical phenomena called photon-assisted tunneling (PAT)~\cite{Kouwenhoven413.2000, Kouwenhoven50.2019}. 
PAT occurs in quantum systems when a photon radiation is applied to an electronic island connected to electron reservoirs~\cite{Shibata109.077401}.
The photon radiation forms extra channels in the electronic island leading to a modification in the electron transport~\cite{Ishibashi314.437}.
Therefore, the photon radiation can play an essential role in the 
transport process generating a photo-current that depends on the photon frequency~\cite{Kouwenhoven73.3443}.
Recently, the influences of photon field, in a vacuum state, 
on two level electronic system~\cite{Guo_Yu_Jie:94205}, and double quantum dots in the presence of a
single mode micro-cavity system with both continuous wave and pulsed excitation are studied~\cite{Ye_Han:114202}. 
Based on the proposed schemes, a single photon generation can be obtained separately under both
QD-cavity resonant and off-resonant conditions. The single photon source, in turn, becomes increasingly important in the 
very diverse range of technological applications.


In addition, external magnetic fields can be used to control the electron transport in nanodevices, 
which leads to several important effective including the change of the energy level spacing inside 
the QD\cite{PhysRevLett.65.108} and hence the QD lowest energy state shrinks with increasing the magnetic field. 
As a result, the Coulomb interaction between two spin degenerate electron grows~\cite{RevModPhys.75.1}.
Furthermore, external magnetic field can form the edge state~\cite{ThomasIhn2010} and the localized state~\cite{PhysRevB.82.195325} 
in the electronic systems, and consequently, the electron transport is reduced.

The combination of the aforementioned fields, namely the magnetic and photon fields, can result 
in a magneto-photon current in graphene~\cite{PhysRevLett.109.267403} and superconductor~\cite{PhysRevB.90.205309}.
In this work, we have considered magneto-photo transport under the influence of a quantized single photon mode in 
a cavity and investigated it's effect on electron transport through a QD system.
In the presence of the photon cavity, extra channels are formed in the system which open new windows 
for electron tunneling called photon-assisted tunneling process.
In addition, we have also shown how an external static magnetic field can control photon-assisted tunneling in the QD system.

This paper is organized as following: In \ref{Model} the description of the model and the theoretical 
formalism are shown. In \ref{Results} we demonstrate the results. 
In the last section, \ref{conclusions}, conclusions will be presented.

\section{Model and Theory}\label{Model}

The system under investigation is a two-dimensional (2D) electron gas exposed to 
an external static magnetic field and a quantized photon field at low temperature. 
We assume that the electronic system consists of a quantum dot embedded in a quantum wire.
The QD system is connected to two electron reservoirs with different chemical potentials.
The electron-photon coupling system is described by the following Hamiltonian in the many-body (MB) basis

\begin{equation}
  \hat{H} = \hat{H}_e + \hat{H}_{\gamma} + \hat{H}_{e\text{-}\gamma},
  \label{eq:H}
\end{equation}
where $H_e$ is the Hamiltonian of the electronic system including electron-electron interaction 
\begin{eqnarray}
 \hat{H}_e &=&\int dr\; \hat{\psi}^{\dagger}(\mathbf{r}) \Big[\frac{\bm{\pi}^2}{2m^*} + \frac{1}{2} 
           m^*\Omega_0^2 y^2 + U_{\rm Dot} + eU_{\rm pg}\Big] \hat{\psi}(\mathbf{r}) \nonumber \\
           & +& \int dr \int dr' \hat{\psi}^{\dagger}(\mathbf{r})  \hat{\psi}^{\dagger}(\mathbf{r}') U_{C}(\mathbf{r},\mathbf{r}') 
           \hat{\psi}(\mathbf{r}') \hat{\psi}(\mathbf{r}).
           \label{eq:H_e}
\end{eqnarray}
Herein, $\bm{\pi} = \mathbf{p}+ (e/c) \mathbf{A}$ with $\mathbf{p}$ and $\mathbf{A} = -By \mathbf{\hat{x}}$
being the canonical momentum and magnetic vector potential, respectively. The magnetic field is applied along the $z$-axis, i.e
$\mathbf{B} = B \mathbf{\hat{z}}$, and $\hat{\psi}(\mathbf{r}) = \sum_i \psi_i(r) d_i$ and
$\hat{\psi}^{\dagger}(\mathbf{r}) = \sum_i \psi_i^*(r) d_i^{\dagger}$ are 
the fermionic field operators with $d_i$($d^{\dagger}_i$) being the annihilation(creation) operators for an 
electron in the single electron state $\ket{i}$ corresponding to $\psi_i$. The QD potential can be described by
\begin{equation}
  U_{\rm Dot} = U e^{(-\alpha_x^2 x^2 - \alpha_y^2 y^2)}.
\end{equation}
Where $U$ is the strength of the potential, and $\alpha_x$ and $\alpha_y$ are constants
that determine the diameter of the QD. The plunger-gate voltage is described by $U_{\rm pg}$ which is 
an electrostatic potential shifting the energy states of the QD system with respect to the chemical potential of 
the leads.
The second term of \eq{eq:H_e} indicates the electron-electron interaction in the central system with 
$U_{\rm C}$ being the Coulomb interaction potential~\cite{Nzar_PhD_Thesis}.

The second part of the \eq{eq:H} can be written as $H_{\gamma} = \hbar \omega_{\gamma} a^{\dagger} a$  
introducing the Hamiltonian of the free photon field 
with $\hbar \omega_{\gamma}$ being the photon energy and $a$($a^{\dagger}$) the photon annihilation(creation) operators.
The quantized vector potential of the cavity photon field, in the Coulomb gauge, is given by 
$\hat{\mathbf{A}}_{\gamma}= A (\hat{a}+\hat{a}^{\dagger}) \mathbf{e}$ where 
$A$ is the amplitude of the photon field, related to the electron-photon coupling constant via $g_{\gamma}=e A a_w \Omega_w/c$, 
and $\mathbf{e}$ determines the photon polarization with either 
parallel $\mathbf{e} = \mathbf{e}_x$ or 
perpendicular  $\mathbf{e} = \mathbf{e}_y$ to the electron motion. Note that $a_w$ is the effective magnetic length and 
$\Omega_w$ is the effective confinement frequency of electrons of the QD system.

The last term on the right side of \eq{eq:H},
\begin{equation}
 \hat{H}_{e\text{-}\gamma} = -\frac{1}{c} \int d{\mathbf{r}} \ \mathbf{j}(\mathbf{r}) \cdot\mathbf{A}_\gamma 
 - \frac{e}{2m^* c^2} \int d{\mathbf{r}} \rho(\mathbf{r}) A_\gamma^2,
\end{equation}
represents the full electron-photon interaction 
including both para- and dia-magnetic electron-photon interactions, respectively. 
The charge is $\rho = -e \psi^{\dagger} \psi$ and the 
charge current density is governed by
\begin{equation}
 \mathbf{j} = -\frac{e}{2m^*}\left\{\psi^\dagger\left({\bm{\pi}}\psi\right)
                 +\left({\bm{\pi}}^*\psi^\dagger\right)\psi\right\}.
\end{equation}
The electron-electron and the electron-photon interactions are treated by exact diagonalization 
in appropriately truncated Fock-spaces.

\begin{figure}[htbq]
\centering
  \includegraphics[width=0.4\textwidth]{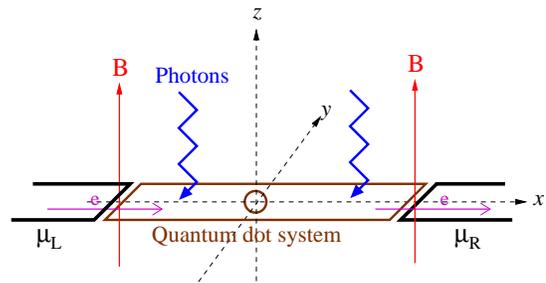}
 \caption{(Color online) Schematic diagram of a quantum dot system (brown color) connected to the left lead (black color) with 
                         chemical potential $\mu_L$ and and the right lead with chemical potential $\mu_R$. The photon field 
                         is represented by the blue zigzag arrows. The external magnetic field $B$ is labeled by a red arrows.}
\label{fig01}
\end{figure}

Figure \ref{fig01} shows the schematic diagram of the quantum dot system (brown color)
connected to two the leads (black color) under the combined effects of the magnetic field $B$ (red arrows)
and the photon radiation (blue zigzag arrows).
The chemical potential of the left lead $\mu_L$ is assumed to be
higher than that of the right lead $\mu_R$. Consequently, the transport is dominated by the left to right 
electron motions between the two leads through 
the central system as indicated by pink arrows.

The Liouville-von Neumann equation is used to describe the time-evolution of the many-body density operator of the closed system.
But in the case of open system when the central system is connected to the leads, we use a projection operator technique 
to derive a generalized master equation for the reduced density operator~\cite{Nakajima20.948,Zwanzing.33.1338}.
Since we are interested in the transient behavior of the system, we assume a non-Markovian approach valid to a weak coupling
of the leads to the central system \cite{PhysRevB.82.195325}. 

Once we have the reduced density operator, one can calculate charge current and charge density in the system.
The charge current is $I^{\rm c}(t) = I^{\rm L}(t) - I^{\rm R}(t)$, where 
$I^{\rm L}(t)$ indicates the partial current from the left lead
into the QD system and $-I^{\rm R}(t)$  refers to the partial
current into the right lead from the QD system.
The partial current can be intoduced as $I^{\rm L,R} = \mathrm{Tr}[\dot{\hat{\rho}}_S^{\rm L,R}(t) \hat{Q}]$ 
where $\dot{\hat{\rho}}_S^{\rm L}$ and $\dot{\hat{\rho}}_S^{\rm R}$ are the time derivatives of 
the system's reduced density matrix due to its coupling to the left and right leads, respectively,
~\cite{Nzar_PhD_Thesis,Nzar_acsphotonics.5b00532} and
$\hat{Q} = e \hat{N}$ is the charge operator with the number operator $\hat{N}$.

\section{Results and Discussions}\label{Results}
We assume the QD system and the leads are made of GaAs semiconductor 
with effective electron mass $m^* = 0.067m_e$ and relative dielectric constant $\kappa = 12.4$.
The parameters of the QD potential are $U = -3.3$~meV, and $\alpha_x = \alpha_y = 0.03$~nm$^{-1}$.
The cavity consists of a single photon mode with energy $\hbar\omega_{\gamma} = 0.3$~meV,
and the electron-photon coupling strength $g_{\gamma} = 0.1$~meV.
The chemical potential of the left and the right leads are $\mu_L = 1.2$~meV and $\mu_R = 1.1$~meV, respectively, 
implying the bias voltage $\Delta \mu = \mu_L - \mu_R = 0.1 $~meV. The temperature of the leads before coupling 
to the QD system is $T = 0.001$~K.
The confinement energy of electrons in the QD system is equal to that of the leads 
$\hbar \Omega_0 = \hbar \Omega_l = 2.0$~meV. Finally, the photon field is linearly polarized and aligned with $x$-axis parallel to
the direction of electron motion in the QD system.

\begin{figure}[htbq!]
       \begin{center}
  \includegraphics[width=0.22\textwidth,angle=0,bb=54 50 215 294,clip]{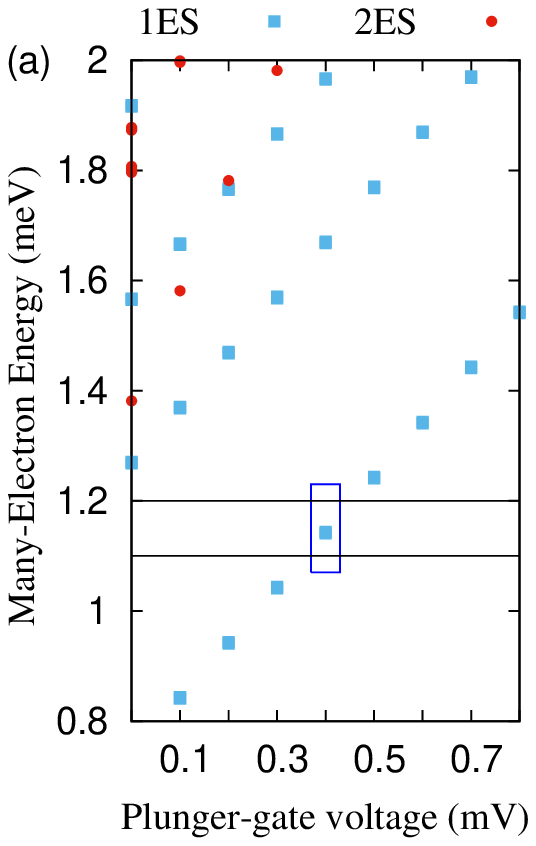}
  \includegraphics[width=0.23\textwidth,angle=0,bb=54 50 215 294,clip]{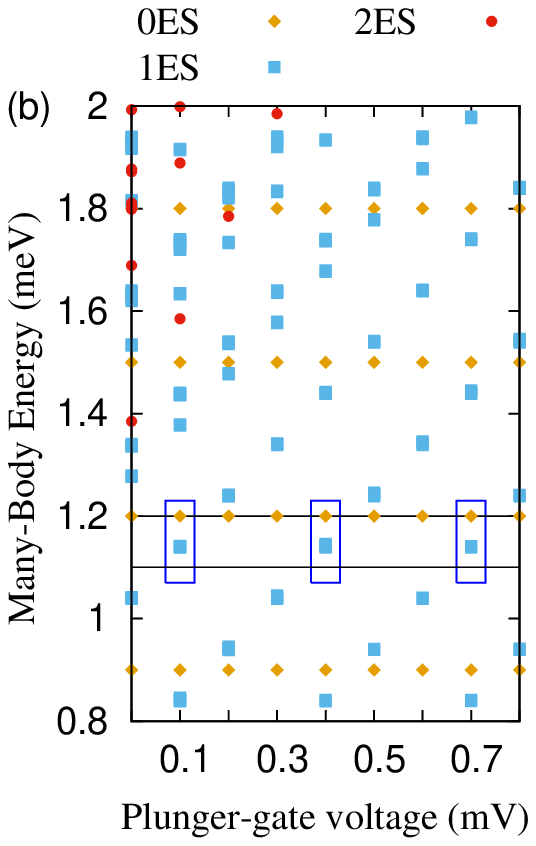}
         \end{center}
 \caption{(Color online) The energy spectrum of the QD system without (a) and with (b) photon cavity 
           versus the plunger-gate voltage $U_{\rm pg}$ including zero-electron states (0ES, golden diamonds), 
           one-electron states (1ES, blue rectangles) and two-electron states (2ES, red dots).
   The chemical potentials are $\mu_L = 1.2\ {\rm meV}$ and $\mu_R = 1.1\ {\rm meV}$ (black). 
The SE state in the bias window is almost doubly degenerate due to the small Zeeman energy.}
       \label{fig02}
\end{figure}

In what follows, we explain the influences of the magnetic field on photon-induced electron transport
through the QD system. Figure \ref{fig02} shows the energy spectrum of the QD system versus the plunger-gate voltage
including zero-electron states (0ES, golden diamonds) one-electron states  (1ES, blue rectangles) and two-electron states (2ES, red circles).
The chemical potential of the leads are indicated by two horizontal lines (black lines).

\begin{figure}[h!]
\centering
  \includegraphics[width=0.5\textwidth,angle=0,bb=54 65 410 294,clip]{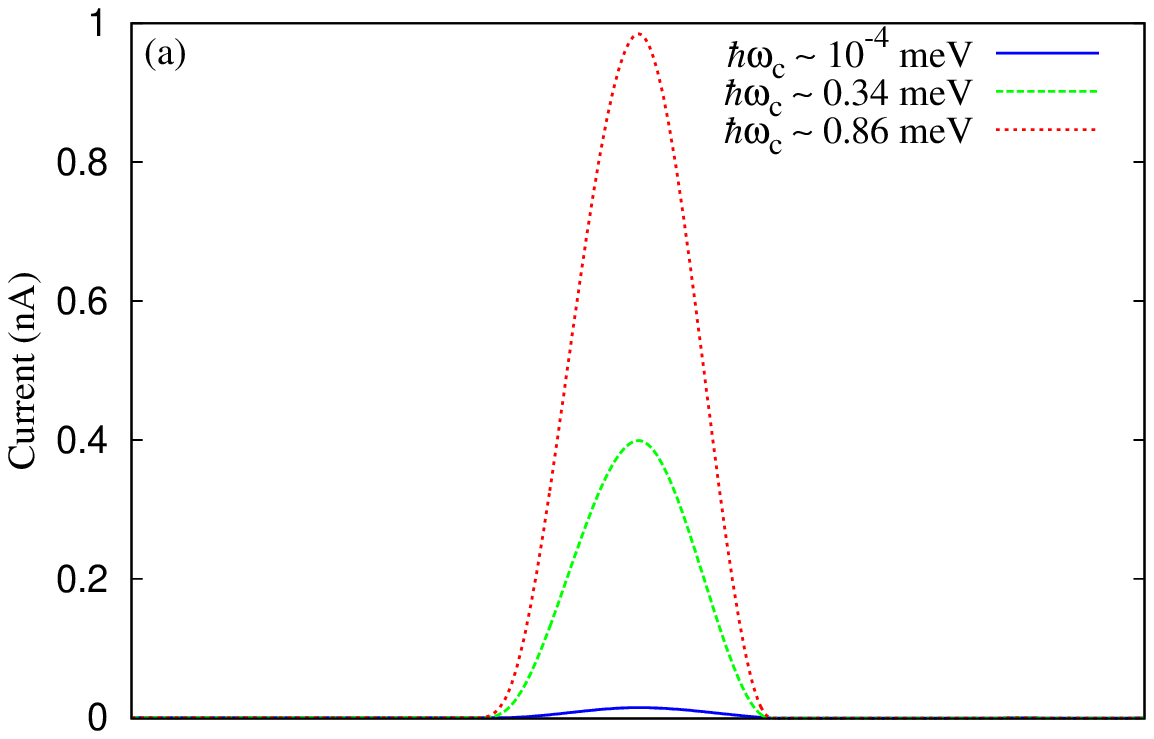}\\
  \includegraphics[width=0.5\textwidth,angle=0,bb=54 50 410 269,clip]{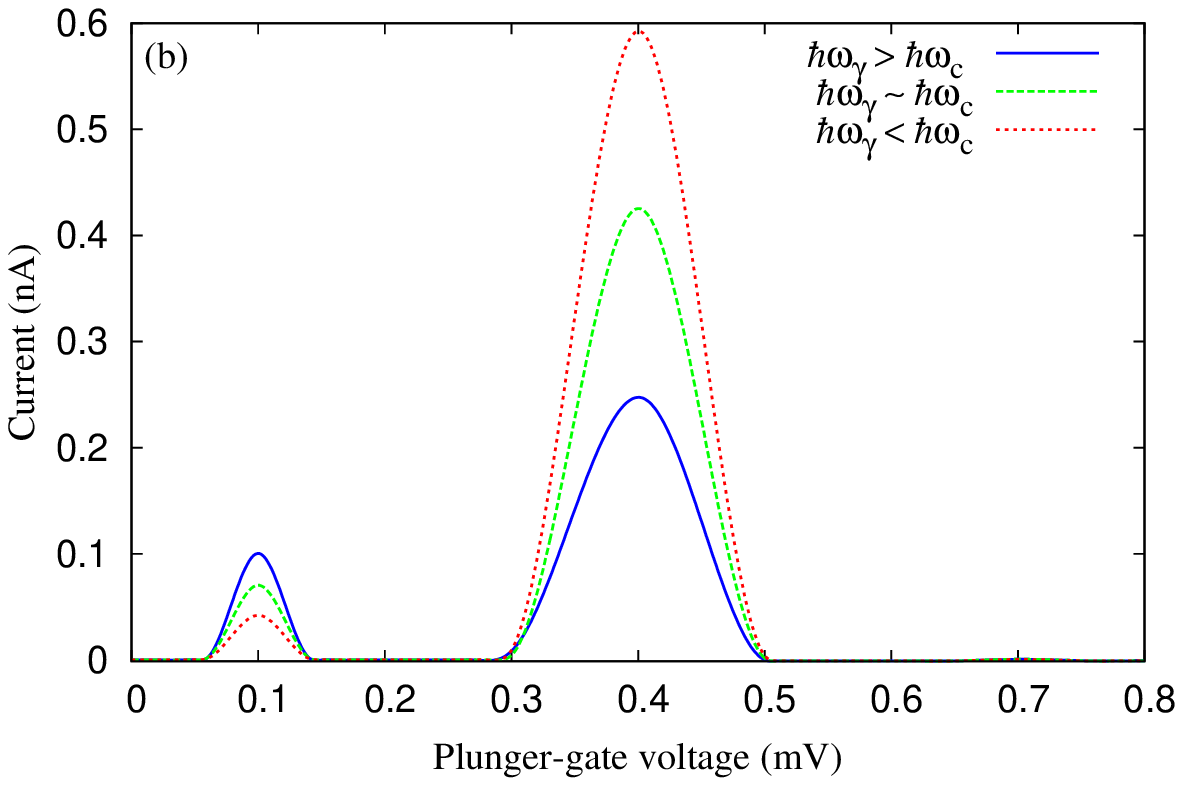}
 \caption{(Color online) The charge current $I_Q$ is plotted as a function of
plunger gate voltage $U_{\rm pg}$ at time $t = 220$~ps for (a) the QD system without photon cavity 
for cyclotron energy $\hbar\omega_c \simeq 10^{-4}$~meV (blue solid), $0.34$~meV  (green dashed) and $0.86$~meV (red dotted),
and (b) the QD system with photon cavity for $\hbar\omega_{\gamma} >  \hbar\omega_c$ (blue solid), 
$\hbar\omega_{\gamma} \simeq \hbar\omega_c$ (green dashed) and $\hbar\omega_{\gamma} < \hbar\omega_c$ (red dotted) 
where the photon energy is $\hbar\omega_{\gamma} =0.3$~meV.
The bias window is $\Delta \mu =0.1~{\rm meV}$, $g_{\gamma} = 0.1$~meV, and $N_{\gamma} = 2$.}
\label{fig03}
\end{figure}

In \fig{fig02}(a) the many-electron (ME) energy of the QD system, excluding the photon cavity, is demonstrated.
For the selected range of the plunger-gate voltage, the first excited state lies between the two chemical potential, inside the bias 
window, which in turn gets into resonance with first subband energy of the leads located in the bias window.
Therefore, an electron in the first subband of the left lead may perform electron tunneling into the 
first-excited state of the QD-system. As a result, a peak in the charge current is formed at $U^0_{\rm pg} = 0.4$~mV 
as is shown in \fig{fig03}. In addition, it should be known that the ground state energy 
of the QD system is found below $0.8$~meV (not shown).

\begin{figure}[h!]
\centering
  \includegraphics[width=0.45\textwidth,angle=0,bb=54 65 410 294,clip]{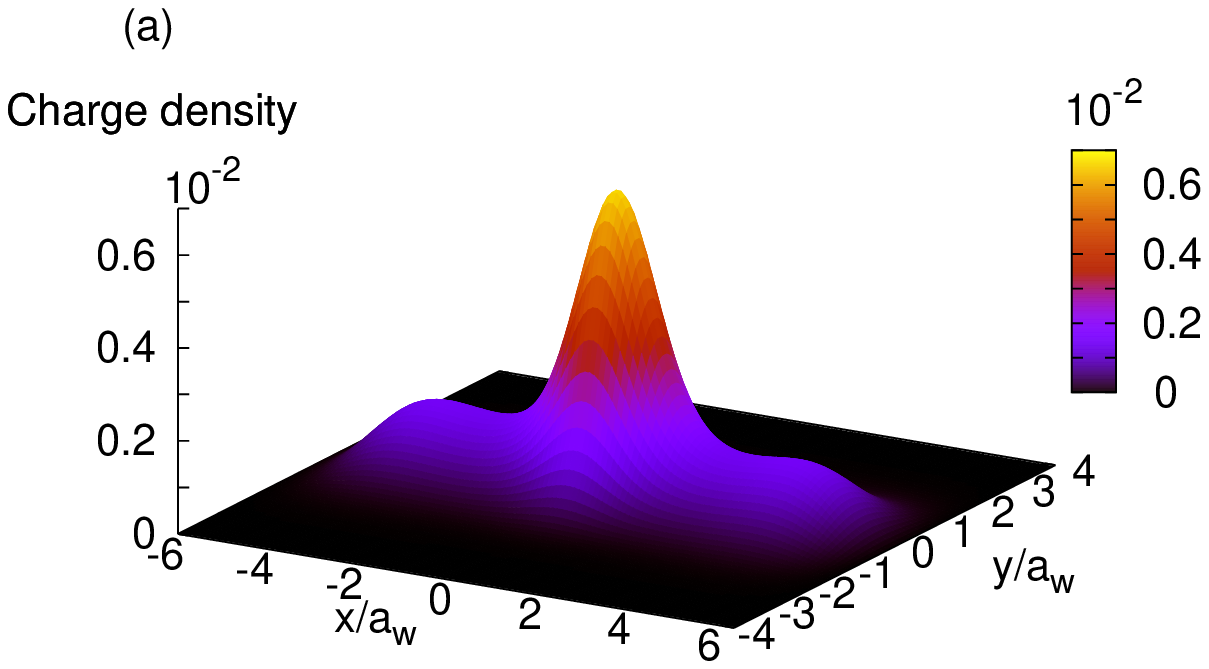}
  \includegraphics[width=0.45\textwidth,angle=0,bb=54 60 410 279,clip]{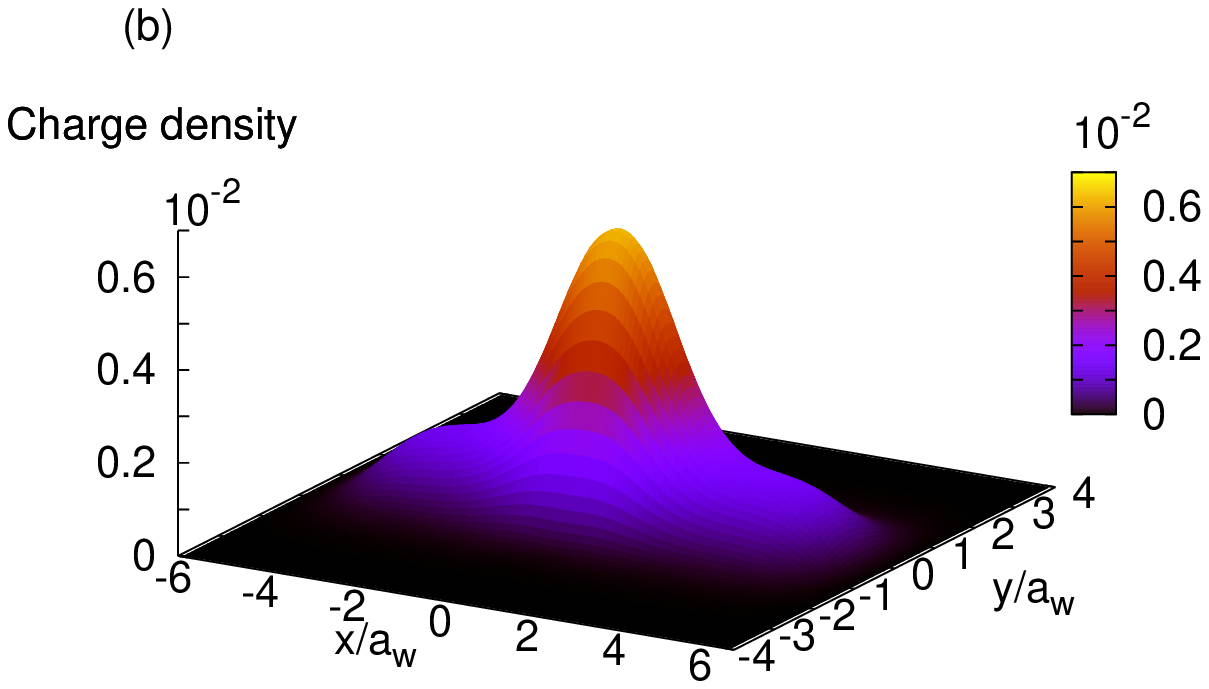}
 \caption{(Color online) Charge density in the QD system at $t = 220$~ps
               without (a) with (b) photon cavity
               in the main current peak at $U^0_{\rm pg} = 0.4$~mV shown in \fig{fig03} 
               for the cyclotron energy $\hbar\omega_c = 10^{-4}$~meV.
               The effective magnetic length is $a_w = 33.72$~nm, $\hbar\omega_{\gamma} = 0.3$~meV, $g_{\gamma} = 0.1$~meV and 
               $N_{\gamma} = 2$.}
\label{fig04}
\end{figure}

In \fig{fig03}(a) the charge current versus the plunger-gate voltage $U_{\rm pg}$ is plotted for 
three different values of the cyclotron energy $\hbar\omega_c \simeq 10^{-4}$~meV (blue solid), $0.34$~meV (green dashed) and 
$0.86$~meV (red dotted) corresponding to the magnetic field $B = 0.0001$~T, $0.2$~T and $0.5$~T, respectively.
We can clearly see that the peak current is increased with the cyclotron energy.
At $\hbar\omega_c \simeq 10^{-4}$~meV the charge current is very weak.
This can be attributed to localization of charge density in the QD (see \fig{fig04}(a)). 
In contrast, for higher value of cyclotron energy (such as $\hbar\omega_c \simeq 0.86$~meV) corresponding 
to the higher magnetic field, 
the charge is delocalized and slightly extended 
to outside of the QD which can be understood as follows.
The high magnetic field induces stronger Lorentz force that forms 
a circular motion of the electron charge density outside the dot (not shown), and 
consequently, the charge current is enhanced (see \fig{fig03}(a) red dotted line).

Now assuming the QD system is coupled to the cavity photon, the electron transport can be 
affected by both the static magnetic and the dynamic photon fields.
For a photon energy  $\hbar\omega_{\gamma} = 0.3$~meV and 
the electron-photon coupling strength $g_{\gamma} = 0.1$~meV, 
\fig{fig02}(b) demonstrates how 
the many-body (MB) energy varies as a function of the plunger-gate voltage.
In the presence of the cavity, photon replica states are formed with different photon contents. 
The energy spacing between the photon replica states is appropriately equal to the photon energy at low electron-photon coupling strength.
Therefore, the first-excited state in the bias window is not active anymore in the electron transport, but instead, 
the electrons from the left leads transfer to the photon replica of the first-excited state in the QD system.
Comparing to the energy spectrum of the QD system in \fig{fig02}(a) for which the photon field is neglected,
in \fig{fig02}(b) two photon replica states at $U_{\rm pg} = 0.1$ and $0.7$~mV (blue rectangles) are found in the bias window
corresponding to $U_{\rm pg} = U_{\rm pg}^0 - \hbar\omega_{\gamma}$ and $U_{\rm pg} = U_{\rm pg}^0 + \hbar\omega_{\gamma}$, respectively.

Figure \ref{fig03}(b) shows the charge current as a function of the plunger-gate voltage in the presence of the photon cavity 
for three cases $\hbar\omega_{\gamma} >  \hbar\omega_c$ (blue solid), $\hbar\omega_{\gamma} \simeq \hbar\omega_c$ (green dashed) 
and $\hbar\omega_{\gamma} < \hbar\omega_c$ (red dotted) where the photon energy is $\hbar\omega_{\gamma} =0.3$~meV.
The peak current (main-peak) at $U_{\rm pg}^0 = 0.4$~meV is again found. In addition to the main peak 
an extra side peak is observed at $U_{\rm pg} = U_{\rm pg}^0 - \hbar\omega_{\gamma}$. The existence of this side peak is due to 
the formation of the one photon replica of the first excited state.

In the case of $\hbar\omega_{\gamma} >  \hbar\omega_c$, where $\hbar\omega_{\gamma} =0.3$~meV and $\hbar\omega_c \simeq 10^{-4}$~meV, 
the photon field is dominant. 
Comparing to the charge current in the absence of the cavity shown in \fig{fig03}(a) (blue solid), 
the current is increased in the main peak at $U^0_{\rm pg} = 0.4$~mV which attributes to the fact that 
the charge density is stretched out of the QD as is shown in \fig{fig04}(b). 
This stretching effect is caused by the paramagnetic term of the electron-photon interaction. 
In a addition, the contribution of photon replica state with two photons can also lead to the 
enhances the charge current. 
It is worth mentioned that this is because the energy of two photon replica state 
is higher than that of the first-excited state in the energy spectrum.
The higher states in the energy spectrum are less bound in the system and actively contribute to the electron transport.

We should also note that the current is almost unchanged when $\hbar\omega_{\gamma} \simeq   \hbar\omega_c$ (green dashed) 
with  $\hbar\omega_c \simeq 0.34$~meV at the main peak 
recapturing the same value of current as it was found in absence photon cavity.

In addition, when $\hbar\omega_{\gamma} < \hbar\omega_c$ (\fig{fig03}(b) red dotted) with 
$\hbar\omega_c \simeq 0.86$~meV the magnetic field effect is dominant.
At this high cyclotron energy the energy spacing between photon replica states are increased and
the photon replica states weakly contribute to the electron transport. As a result, 
the charge current is decreased in the main peak.

\begin{figure}[h!]
\centering
  \includegraphics[width=0.45\textwidth,angle=0,bb=54 65 410 294,clip]{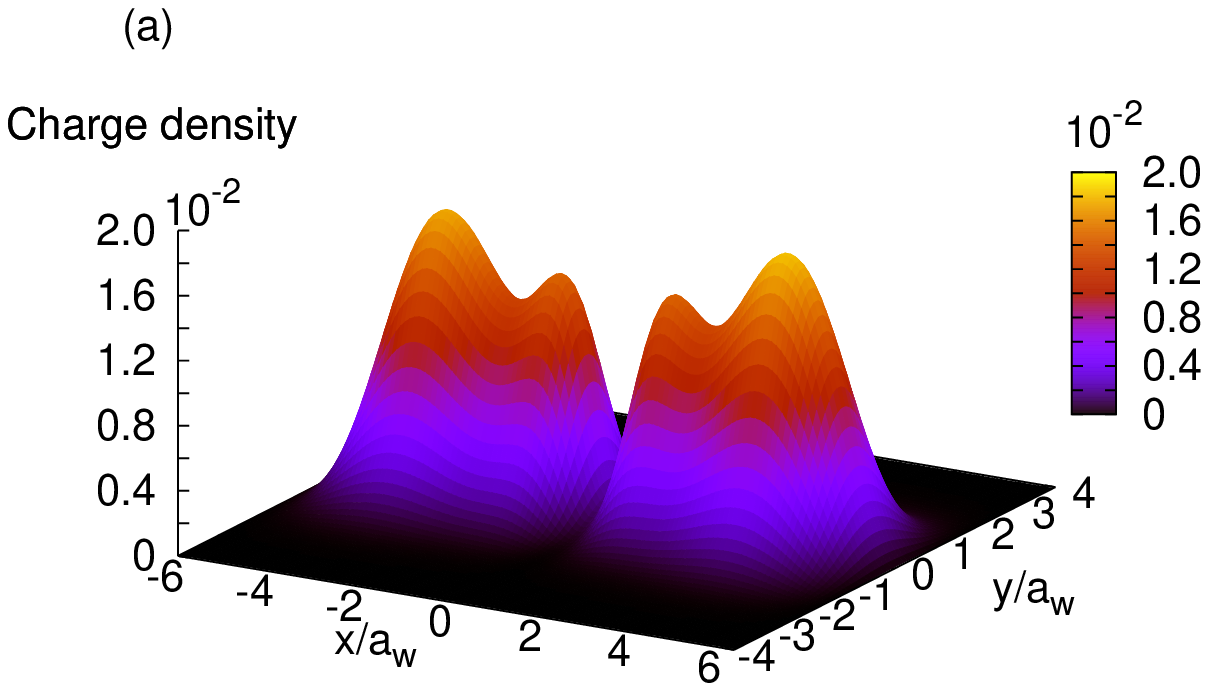}
  \includegraphics[width=0.45\textwidth,angle=0,bb=54 60 410 279,clip]{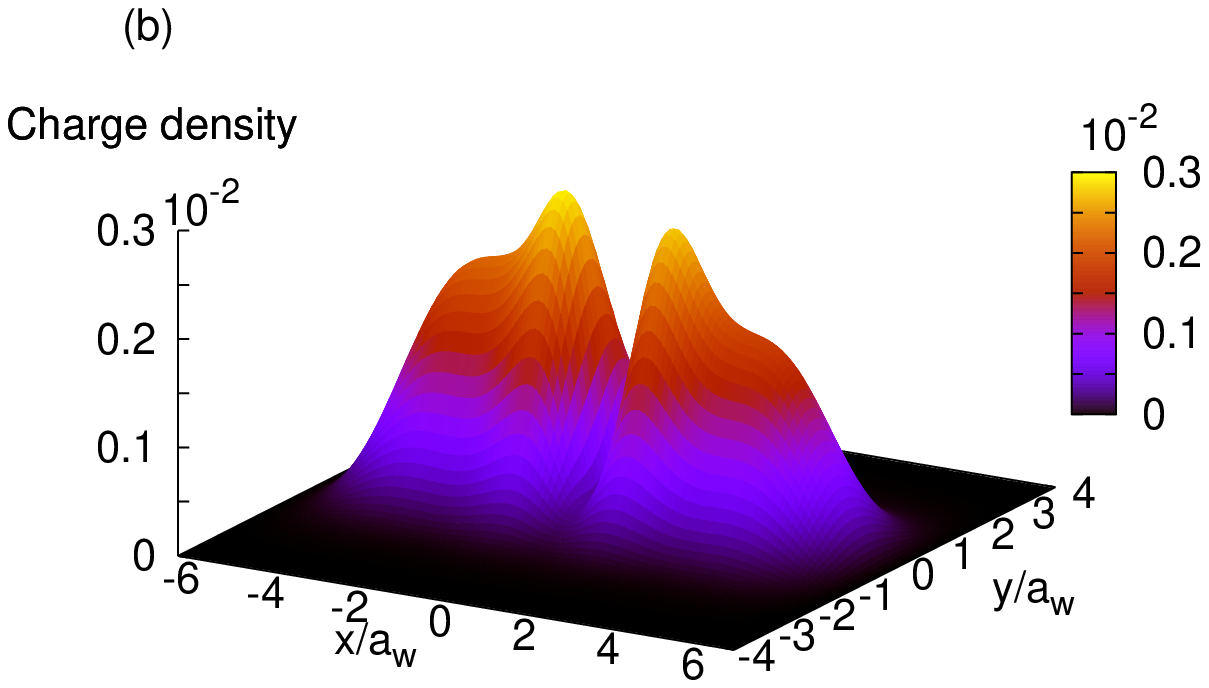}
 \caption{(Color online) Charge density in the QD system at $t = 220$~ps
               with photon cavity in the side current peak at $U_{\rm pg} = 0.1$~mV shown in \fig{fig03}(b) 
               for case of $\hbar\omega_{\gamma} > \hbar\omega_c$ (a) and $\hbar\omega_{\gamma} < \hbar\omega_c$ (b).
               The effective magnetic length is $a_w = 33.72$~nm, $\hbar\omega_{\gamma} = 0.3$~meV, $g_{\gamma} = 0.1$~meV and 
               $N_{\gamma} = 2$.}
\label{fig05}
\end{figure}

Another interesting aspect of this issue is the influence of the external magnetic field on the current in the side peak displayed in 
\fig{fig03}(b). The formation of side peak is totally due to the photon cavity.  
It can be clearly seen that the current is high at low cyclotron energy when $\hbar\omega_{\gamma} > \hbar\omega_c$ (blue solid)
indicating that the photon-induced current should be generated at low magnetic field. 
There are two reasons for the high current here: the photon replica states are more active in the electron transport 
at low cyclotron energy, and the stretching of charge density in the QD system due to the photon cavity.
To explain the enhancement of curent at the side peak, the charge density at $U_{\rm pg} = 0.1$~mV
is shown in \fig{fig05} for (a) the low cyclotron energy ($\hbar\omega_c \simeq 10^{-4}$~meV), i.e $\hbar\omega_{\gamma} > \hbar\omega_c$, 
and (b) the high cyclotron energy ($\hbar\omega_c \simeq 0.86$~meV), i.e $\hbar\omega_{\gamma} < \hbar\omega_c$.
In \fig{fig05}(a) the charge density is mostly distributed outside the QD 
and near the contact area to the leads. The charge accumulation in the contact area leads to the 
stronger charging to the QD system from the leads, and then the charge current at the side peak 
is increased. Therefore, we emphasis that the PAT process requires the following condition $\hbar\omega_{\gamma} > \hbar\omega_c$.

By contrast, at high cyclotron energy ($\hbar\omega_c \simeq 0.86$~meV) 
when $\hbar\omega_{\gamma} < \hbar\omega_c$, the magnetic field dominates the electron transport.
The magnetic filed causes the charge accumulation around the QD as is shown in \fig{fig05}(b), and then
the current is reduced at the side peak.

\section{Conclusions}\label{conclusions}

We have investigated the influences of a static magnetic field on photon-induced transport
through a quantum dot coupled to a quantized photon cavity.
It was found that the cavity forms photon replica states and their contribution to the electron transport
can be affected by the external magnetic field. 
Therefore, two different regimes are studied, low and high magnetic fields.  
At the low magnetic field regime (low cyclotron energy) where the cyclotron energy is assuemd to be lower than the 
photon energy, the photon replica states formed in the presence of the cavity 
are actively contribute to the transport. Consequently, the charge density in the system are stretched and
then the current is increased. 

On the other hand, at the high magnetic field regime (high cyclotron energy), 
when the cyclotron energy is higher than the photon energy, the magnetic field is dominant
and the photon-induced current is suppressed. 
As a result, we emphasis that the photon-induced transport or photon-assisted transport can be obtained when the 
the photon energy is higher than the cyclotron energy in the system.

\begin{acknowledgments}
 Financial support is acknowledged from the Icelandic Research and Instruments Funds, 
and the Research Fund of the University of Iceland. The calculations were carried out on 
the Nordic High Performance Computer Center in Iceland. We are thankful to Dr. Peshwaz Abdoul
for very interesting discussions. We acknowledge the Nordic network
NANOCONTROL, and The University of Sulaimani. 
\end{acknowledgments}

%

\begin{thebibliography}{22}

\bibitem{Imamog72.210}
A.~Imamoglu and Y.~Yamamoto.
\newblock {Turnstile device for heralded single photons: Coulomb blockade of
  electron and hole tunneling in quantum confined \textit{p} - \textit{i} -
  \textit{n} heterojunctions}.
\newblock {\em Phys. Rev. Lett.}, 72:210--213, Jan 1994.

\bibitem{Loss.57.1998}
D.~Loss and D.~P. DiVincenzo.
\newblock {Quantum computation with qunatum dots}.
\newblock {\em Physical Review A}, 57(1):120, 1998.

\bibitem{DiVincenzo.309.2005}
D.~P. DiVincenzo.
\newblock {Double quantum dot as a quantum bit}.
\newblock {\em Science}, 309, 2173 2005.

\bibitem{Petroff1997}
Pierre~M. Petroff, Klaus~H. Schmidt, Gilberto~Medeiros Ribeiro, Axel Lorke, and
  Jorg Kotthaus.
\newblock {Size Quantization and Zero Dimensional Effects in Self Assembled
  Semiconductor Quantum Dots}.
\newblock {\em Japanese Journal of Applied Physics}, 36(6S):4068, 1997.

\bibitem{Kouwenhoven1999}
LeoP. Kouwenhoven and PaulL. McEuen.
\newblock {Single Electron Transport Through a Quantum Dot}.
\newblock In Gregory Timp, editor, {\em {Nanotechnology}}, pages 471--535.
  Springer New York.

\bibitem{Kouwenhoven413.2000}
T.~Fujisawa, W.~G. van~der Wiel, and L.~P. Kouwenhoven.
\newblock {Inelastic tunneling in a double quantum dot coupled to a bosonic
  enviroment}.
\newblock {\em Physica E}, 7:413--419, 2000.

\bibitem{Kouwenhoven50.2019}
L.~P. Kouwenhoven, S.~Jauhar, K.~McCormick, D.~Dixon, P.~L. McEuen, Yu.~V.
  Nazarov, N.~C. van~der Vaart, and C.~T. Foxon.
\newblock {Photon-assisted tunneling through a quantum dot}.
\newblock {\em Phys. Rev. B}, 50:2019--2022, Jul 1994.

\bibitem{Shibata109.077401}
K.~Shibata, A.~Umeno, K.~M. Cha, and K.~Hirakawa.
\newblock {Photon-Assisted Tunneling through Self-Assembled InAs Quantum Dots
  in the Terahertz Frequency Range}.
\newblock {\em Phys. Rev. Lett.}, 109:077401, Aug 2012.

\bibitem{Ishibashi314.437}
K~Ishibashi and Y~Aoyagi.
\newblock {Interaction of electromagnetic wave with quantum dots}.
\newblock {\em Physica B}, 314:437--443, 2002.

\bibitem{Kouwenhoven73.3443}
L.~P. Kouwenhoven, S.~Jauhar, J.~Orenstein, P.~L. McEuen, Y.~Nagamune,
  J.~Motohisa, and H.~Sakaki.
\newblock {Observation of Photon-Assisted Tunneling through a Quantum Dot}.
\newblock {\em Phys. Rev. Lett.}, 73:3443--3446, Dec 1994.

\bibitem{Guo_Yu_Jie:94205}
Nie Wen-Jie {Guo Yu-Jie}.
\newblock {\em Chinese Physics B}, 24(9):94205, 2015.

\bibitem{Ye_Han:114202}
Yu~Zhong-Yuan Zhang Wen Liu Yu-Min {Ye Han}, Peng Yi-Wei.
\newblock {Sub-Poissonian photon emission in coupled double quantum
  dots\&\#8211;cavity system}.
\newblock {\em Chinese Physics B}, 24(11):114202, 2015.

\bibitem{PhysRevLett.65.108}
P.~A. Maksym and Tapash Chakraborty.
\newblock {Quantum dots in a magnetic field: Role of electron-electron
  interactions}.
\newblock {\em Phys. Rev. Lett.}, 65:108--111, Jul 1990.

\bibitem{RevModPhys.75.1}
W.~G. van~der Wiel, S.~{De Franceschi}, J.~M. Elzerman, T.~Fujisawa,
  S.~Tarucha, and L.~P. Kouwenhoven.
\newblock {Electron transport through double quantum dots}.
\newblock {\em Rev. Mod. Phys.}, 75:1--22, Dec 2002.

\bibitem{ThomasIhn2010}
Thomas Ihn.
\newblock {\em {Semiconductor Nanostructures}}.
\newblock Oxford University Press, New York, US, 2010.

\bibitem{PhysRevB.82.195325}
Nzar~Rauf Abdullah, Chi-Shung Tang, and Vidar Gudmundsson.
\newblock {Time-dependent magnetotransport in an interacting double quantum
  wire with window coupling}.
\newblock {\em Phys. Rev. B}, 82:195325, Nov 2010.

\bibitem{PhysRevLett.109.267403}
David Hagenm\"{u}ller and Cristiano Ciuti.
\newblock {Cavity QED of the Graphene Cyclotron Transition}.
\newblock {\em Phys. Rev. Lett.}, 109:267403, Dec 2012.

\bibitem{PhysRevB.90.205309}
Curdin Maissen, Giacomo Scalari, Federico Valmorra, Mattias Beck,
  J\'{e}r\^{o}me Faist, Sara Cibella, Roberto Leoni, Christian Reichl,
  Christophe Charpentier, and Werner Wegscheider.
\newblock {Ultrastrong coupling in the near field of complementary split-ring
  resonators}.
\newblock {\em Phys. Rev. B}, 90:205309, Nov 2014.

\bibitem{Nzar_PhD_Thesis}
N.~R. Abdullah.
\newblock {\em {Cavity-photon controlled electron transport through quantum
  dots and waveguide systems}}.
\newblock PhD Thesis, University of Iceland, Reykjavik, Iceland (2015).

\bibitem{Nakajima20.948}
S.~Nakajima.
\newblock {On quantum theory of transport phenomena steady diffusion}.
\newblock {\em Prog. of Theor. Phys.}, 20:948, 1958.

\bibitem{Zwanzing.33.1338}
Robert Zwanzig.
\newblock {Ensemble Method in the Theory of Irreversibility}.
\newblock {\em The Journal of Chemical Physics}, 33(5):1338--1341, 1960.

\bibitem{Nzar_acsphotonics.5b00532}
Nzar~Rauf Abdullah, Chi-Shung Tang, Andrei Manolescu, and Vidar Gudmundsson.
\newblock {\em ACS Photonics}, 3(2):249--254, 2016.

\end{thebibliography}
%

%
\end{document}